\documentclass[pra, twocolumn, showpacs, showkeys]{revtex4}

\usepackage{amsmath}
\usepackage{amssymb}
\usepackage[dvips]{graphicx}

\DeclareGraphicsExtensions{.ps}

\newcommand*{\etal}{\textit{et al.\ }}

\begin{document}
\title{Transmission of ultracold atoms through a micromaser~: detuning effects}
\date{17 February 2004}
\author{John Martin}
\email{John.Martin@ulg.ac.be}
\author{Thierry Bastin}
\email{T.Bastin@ulg.ac.be}
 \affiliation{Institut de Physique
Nucl\'eaire, Atomique et de Spectroscopie, Universit\'e de Li\`ege
au Sart Tilman, B\^at.\ B15, B - 4000 Li\`ege, Belgique}

\begin{abstract}
The transmission probability of ultracold atoms through a
micromaser is studied in the general case where a detuning between
the cavity mode and the atomic transition frequencies is present.
We generalize previous results established in the resonant case
(zero detuning) for the mesa mode function. In particular, it is
shown that the velocity selection of cold atoms passing through
the micromaser can be very easily tuned and enhanced using a
non-resonant field inside the cavity. Also, the transmission
probability exhibits with respect to the detuning very sharp
resonances that could define single cavity devices for high
accuracy metrology purposes (atomic clocks).
\end{abstract}


\maketitle

\section{Introduction}
\label{Introduction}

Laser cooling of atoms has become these last years an interesting
tool in atom optics (see e.g.\ Refs.~\cite{Ada94,Ada97} for a
review of these topics). The production of slow atomic beams and
the control of their motion by laser light has opened a variety of
applications including matter-wave interferometers \cite{Kas91},
atomic lenses or atom lithography \cite{Tim92}. In such
experiments, it is often highly desirable to control actively the
velocity distribution of an atomic ensemble. More particularly,
devices for narrowing this distribution over a well fixed velocity
are very useful to define an atomic beam with a long coherence
length (like in atom lasers \cite{Ket02}). Velocity
monochromatization of atomic beams making use of an optical cavity
has already been suggested by Balykin \cite{Bal89}. For ultracold
atoms, L\"{o}ffler \etal \cite{Lof98} have recently proposed a
velocity selector based on a 1D micromaser scheme (also referred
to as mazer). They suggest to send a beam of cold atoms through a
microwave cavity in resonance with one of the atomic transitions.
The small velocity of the atoms at the entrance of the micromaser
require to quantize their center-of-mass motion to describe
correctly their interaction with the cavity quantum field (see
e.g.\ Refs.~\cite{Sch97a,Bas03b} for an overview of quantized
motion in quantized fields). This quantization is an essential
feature as it leads to a fundamental interplay between their
motion and the atom-field internal state \cite{Scu96}. It results
from this that most of the incoming atoms may be found reflected
by the field present in the cavity, except at certain velocities
where they can be transmitted through with a reasonable
efficiency. At the exit of the cavity, the longitudinal velocity
distribution of the cold atomic beam may be this way significantly
narrowed \cite{Lof98} and a splitting of the atomic wave packet
may be observed \cite{Bie01}. These effects have been first
described by Haroche \etal \cite{Har91}, Englert \etal
\cite{Eng91} and Battocletti and Englert \cite{Bat94}. They have
been recently experimentally observed in the optical domain (see,
for example, Pinkse \etal \cite{Pin00}).

In this paper, we extend the proposal of L\"{o}ffler \etal
\cite{Lof98} by considering an off-resonant interaction between
the atoms and the cavity field. This case offers new attractive
perspectives for metrology purposes and in the velocity selection
scheme. Let us emphasize that this scheme is in no way a proposal
to reduce the transverse momentum spread of an atomic beam. For
applications where this point is important (like for example in
the case of an atomic beam splitter based on Doppleron resonances
\cite{Gla91}), other schemes like quantum-nondemolition
measurement of atomic momentum \cite{Sle93} should rather be used.

\section{Transmission probability through the mazer}

We consider two-level atoms moving along the $z$ direction on the
way to a cavity of length $L$. The atoms are coupled
off-resonantly to a single mode of the quantized field present in
the cavity. The atomic center-of-mass motion is described quantum
mechanically and the usual rotating-wave approximation is made.
The Hamiltonian of the system reads
\begin{equation}
    \label{Hamiltonian}
        H = \hbar \omega_0 \sigma^{\dagger} \sigma + \hbar \omega a^{\dagger} a + \frac{p^2}{2m}+ \hbar g  u(z) (a^{\dagger} \sigma + a
        \sigma^{\dagger}),
\end{equation}
where $p$ is the atomic center-of-mass momentum along the $z$
axis, $m$ the atomic mass, $\omega_0$ the atomic transition
frequency, $\omega$ the cavity field mode frequency, $\sigma = |b
\rangle \langle a|$ ($|a\rangle$ and $|b\rangle$ are respectively
the upper and lower levels of the two-level atom), $a$ and
$a^{\dagger}$ are respectively the annihilation and creation
operators of the cavity radiation field, $g$ is the atom-field
coupling strength and $u(z)$ is the cavity field mode. We denote
also hereafter $\kappa = \sqrt{2mg/\hbar}$, $\kappa_n = \kappa
\sqrt[4]{n + 1}$, $\delta$ the detuning $\omega - \omega_0$, and
$\theta_n$ the angle defining the dressed-state basis given by
\begin{equation}
    \label{thetads}
    \cot 2 \theta_n = - \frac{\delta}{\Omega_n}\, ,
\end{equation}
with $\Omega_n = 2 g \sqrt{n + 1}$.

The properties of the mazer have been established in the resonant
case by Scully and collaborators \cite{Mey97,Lof97,Sch97}. We
extended very recently these studies in the non-resonant case
\cite{Bas03b}, especially for the mesa mode function ($u(z) = 1$
inside the cavity, 0 elsewhere). Particularly, we have shown that,
if the cavity field is prepared in the Fock state $|n\rangle$, an
atom initially in the excited state $|a\rangle$ with a momentum
$\hbar k$ will be found transmitted by the cavity in the same
state or in the lower state $|b\rangle$ with the respective
probabilities
\begin{equation}
    \label{Tan}
    T^a_n(k) = |\tau^a_n(k)|^2,
\end{equation}
and
\begin{equation}
    \label{Tbnpone}
    \qquad T^b_{n+1}(k) = \left\{ \begin{array}{ll}
    \frac{k_b}{k}|\tau^b_{n+1}(k)|^2 & \textrm{ if  } \left(\frac{k}{\kappa}\right)^2 >
    \frac{\delta}{g},
    \vspace{0.2cm} \\ 0 & \textrm{ otherwise},
    \end{array} \right.
\end{equation}
where
\begin{equation}\label{cab}
k^2_b=k^2-\kappa^2 \frac{\delta}{g},
\end{equation}
and
\begin{align}
    \tau^a_n(k) & = \frac{\cos^2 \theta_n \frac{\tau^-_n(k)}{\tau^-_n(k_b)} \, \tau^+_n(k_b) + \sin^2 \theta_n \, {\tau^-_n(k)}}{\left( \cos^2 \theta_n \frac{k - k_b}{k^c_n} -1 \right) \left( \cos^2 \theta_n \frac{k - k_b}{k^t_n} -1 \right)}, \\
    \tau^b_{n + 1}(k) & = \frac{\sin 2 \theta_n}{4} \left( 1 + \frac{k}{k_b}
    \right)\nonumber\\
    & \;\;\times\frac{\frac{\tau^-_n(k)}{\tilde{\tau}^-_n(k,k_b)} \, \tau^+_n(k_b) - \frac{\tau^+_n(k_b)}{\tilde{\tau}^+_n(k,k_b)} \, \tau^-_n(k)}{\left( \cos^2 \theta_n \frac{k - k_b}{k^c_n} -1 \right) \left( \cos^2 \theta_n \frac{k - k_b}{k^t_n} -1
    \right)},
\end{align}
with
\begin{equation}
\tau^{\pm}_n(k)=\left[ \cos(k^{\pm}_n L) - i \Sigma^{\pm}_n(k)
\sin(k^{\pm}_n L) \right]^{-1},
\end{equation}
\begin{equation}
\tilde{\tau}^{\pm}_n(k,k_b)=\left[ \cos(k^{\pm}_n L) - i
\tilde{\Sigma}^{\pm}_n(k, k_b) \sin(k^{\pm}_n L) \right]^{-1},
\end{equation}
\begin{align}
    {k^+_n}^2 & = k^2 - \kappa_n^2 \tan \theta_n, \\
    {k^-_n}^2 & = k^2 + \kappa_n^2 \cot \theta_n,
\end{align}
\begin{equation}
\Sigma^{\pm}_n(k) = \frac{1}{2}\left( \frac{k^{\pm}_n}{k} +
\frac{k}{k^{\pm}_n} \right),
\end{equation}
\begin{equation}
\tilde{\Sigma}^{\pm}_n(k, k_b) = \left( \frac{k^{\pm}_n}{k + k_b}
+ \frac{k_b}{k + k_b} \frac{k}{k^{\pm}_n} \right),
\end{equation}
\begin{equation}
k^c_n = i \frac{\left(k + i \cot(\frac{k^-_n L}{2}) k^-_n \right)
\left( k_b + i \cot(\frac{k^+_n L}{2}) k^+_n\right)}{
\cot(\frac{k^-_n L}{2}) k^-_n - \cot(\frac{k^+_n L}{2}) k^+_n},
\end{equation}
\begin{equation}
k^t_n = i \frac{\left(k - i \tan(\frac{k^-_n L}{2}) k^-_n \right)
\left( k_b - i \tan(\frac{k^+_n L}{2})
k^+_n\right)}{\tan(\frac{k^+_n L}{2}) k^+_n - \tan(\frac{k^-_n
L}{2}) k^-_n}.
\end{equation}

The atom transmission in the lower state $|b\rangle$ results in a
photon induced emission inside the cavity. In presence of a
detuning, these atoms are found to propagate with a momentum
$\hbar k_b$ different from the initial value $\hbar k$ (see
\cite{Bas03b}). This results merely from the energy conservation.
Contrary to the resonant case, the final state of the process $|b,
n+1\rangle$ has an internal energy different from that of the
initial one ($|a, n\rangle$). The energy difference $\hbar \delta$
is transferred to the atomic kinetic energy. According to the sign
of the detuning, the atoms are either accelerated ($\delta < 0$,
heating process) or decelerated ($\delta > 0$, cooling process).
In this last case, the initial atomic kinetic energy ($\hbar^2
k^2/2 m$) must be greater than $\hbar \delta$ to ensure that the
photon emission may occur. This justifies the conditional result
in Eq.~(\ref{Tbnpone}).

In the ultracold regime ($k \ll \kappa_n \sqrt{\tan \theta_n}$)
and for $\exp(\kappa_n L) \gg 1$ we have $\tau^+_n(k) \simeq 0$
and the total transmission probability $T_n(k) = T^a_n(k) + T^b_{n
+ 1}(k)$ simplifies to
\begin{equation}
    \label{TransmissionT}
    \qquad T_n(k) = f(\theta_n) \mathcal{I}(L) |\tau^-_n(k)|^2
\end{equation}
with
\begin{equation}
    f(\theta_n) = \left\{
    \begin{array}{ll}
            \sin^2 \theta_n ( \sin^2 \theta_n + \frac{k_b}{k} \cos^2 \theta_n )
          & \textrm{ if  } \left(\frac{k}{\kappa}\right)^2 > \frac{\delta}{g}, \\
            \sin^4 \theta_n
          & \textrm{ otherwise}.
    \end{array}\right.
\end{equation}
\begin{equation}
    \mathcal{I}(L) = \frac{1}{\left| \cos^2 \theta_n \frac{k - k_b}{k^c_n} - 1 \right|^2 \left| \cos^2 \theta_n \frac{k - k_b}{k^t_n} - 1
    \right|^2},
\end{equation}
and
\begin{equation}
    |\tau^-_n(k)|^2 \simeq \frac{1}{1 + \left(\frac{\kappa_n}{2k}\right)^2 \cot \theta_n \sin^2(k_n^-
    L)}.
\end{equation}

At resonance ($\delta = 0$), $k_b = k$, $\theta_n = \pi/4$ and
Eq.~(\ref{TransmissionT}) well reduces to the result of L\"offler
\etal \cite{Lof98}
\begin{equation}
    T_n(k) = \frac{1}{2}|\tau^-_n(k)|^2 = \frac{1}{2}\frac{1}{1 + \left(\frac{\kappa_n}{2k}\right)^2 \sin^2(k_n^-
    L)}.
\end{equation}

\begin{figure*}
\includegraphics[width=8.5cm, bb=113 284 475
525]{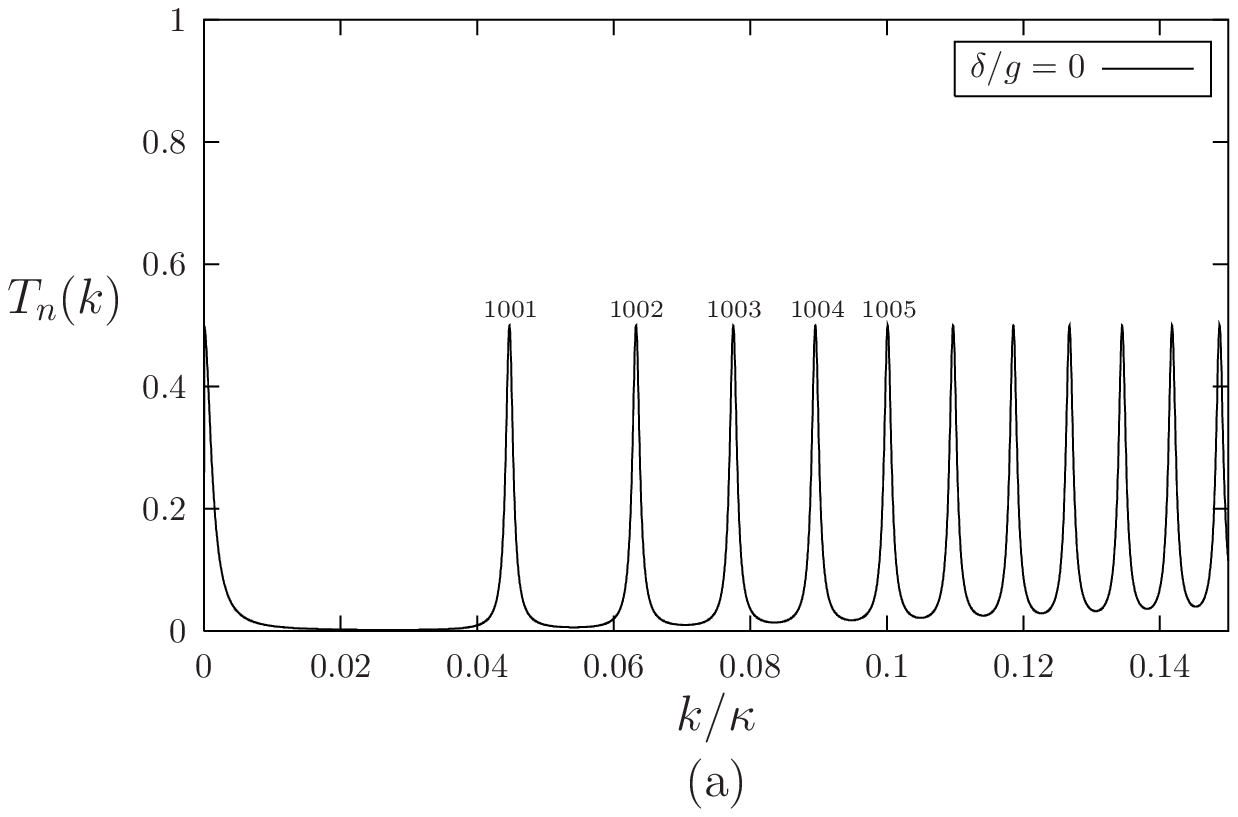}\hspace{5pt}
\includegraphics[width=8.5cm, bb=113 284 475 525]{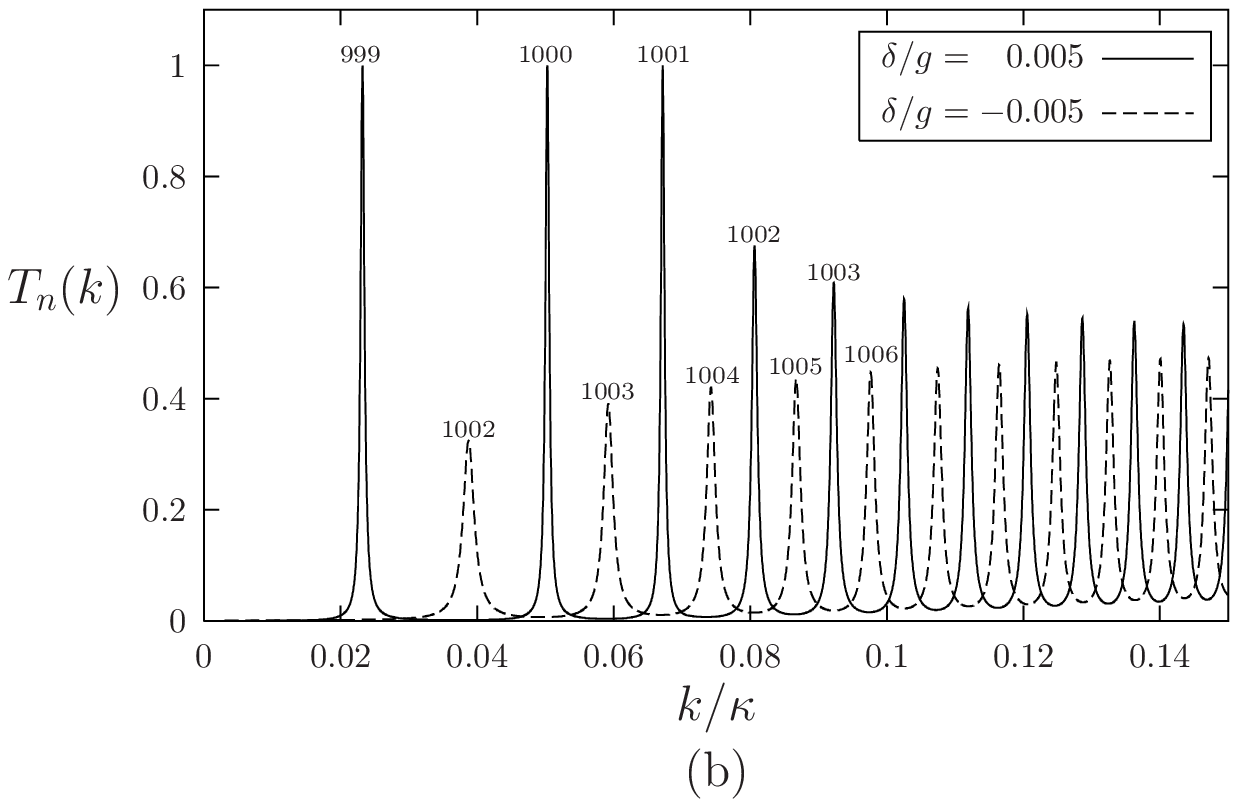}
\caption{Transmission probability of an excited atom through the
mazer with respect to $k/\kappa$. (a) $\delta/g = 0$, (b)
$\delta/g = \pm 0.005$. The interaction length was fixed to
$\kappa L = 10^3 \pi$ and $n = 0$.} \label{TFigs}
\end{figure*}

We present on figs.~\ref{TFigs} the transmission probability of an
initially excited atom through the mazer. With respect to the
wavenumber of the incoming atoms $k/\kappa_n$ or the interaction
length $\kappa_n L$, the transmission probability $T$ shows
various resonances. For $(k/\kappa)^2 > \delta/g$, their position
is given by
\begin{equation}
    \label{peakposition}
    k_n^- L = m \pi \quad (m
    \textrm{ a positive integer}).
\end{equation}

As the de Broglie wavelength is given in the ultracold regime by
$\lambda_{\mathrm{dB}}=2\pi/k^-_{n}$, this occurs when the cavity
length fits a multiple of half the de Broglie wavelength
$\lambda_{\mathrm{dB}}$ of the atom inside the cavity~:
\begin{equation}
    \label{LeqldB}
    L = m \frac{\lambda_{\mathrm{dB}}}{2}.
\end{equation}

The position of the $m^{\textrm{th}}$ resonance in the $k$ space
is therefore given by
\begin{equation}\label{reso}
  \left.\frac{k}{\kappa}\right|_m=\sqrt{\left(\frac{m\pi}{\kappa
  L}\right)^2-\sqrt{n+1}\cot\theta_n}.
\end{equation}

For $(k/\kappa)^2 \leq \delta/g$, a careful analysis of the
transmission probability (\ref{TransmissionT}) yields resonance
positions slightly shifted from the values given by
Eq.~(\ref{peakposition}).

\begin{figure}[htb]
\includegraphics[width=8.5cm, bb=113 284 475 525]{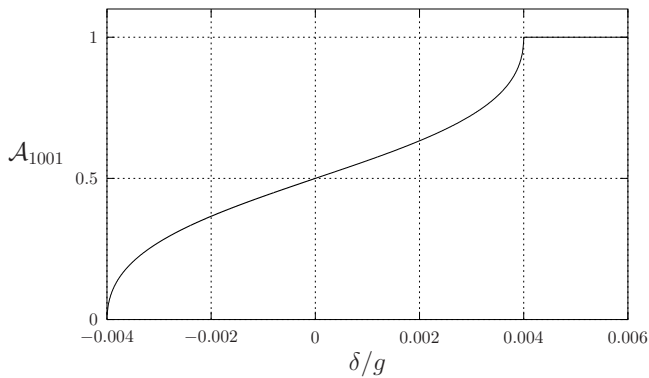}
\caption{Amplitude of the $1001^{\textrm{st}}$ resonance with
respect to the detuning value ($n = 0$, $\kappa L = 10^3 \pi$).}
\label{AFig}
\end{figure}

We have labelled most resonances of figs.~\ref{TFigs} by the
corresponding integer $m$. The amplitude $\mathcal{A}_m$ of a
given resonance strongly depends on the detuning value (see
fig.~\ref{AFig}). We have
\begin{equation}
    \mathcal{A}_m = \left\{
    \begin{array}{ll}
            \displaystyle f(\theta_n) \mathcal{I}(m \frac{\lambda_{\mathrm{dB}}}{2}) \simeq \frac{4 f(\theta_n)}{(1 + \left.\frac{k_b}{k}\right|_{m})^2}
          & \displaystyle \textrm{ if  } \left.\frac{k}{\kappa}\right|_m > \sqrt{\frac{\delta}{g}}, \vspace{0.2cm} \\
            1
          & \textrm{ otherwise}.
    \end{array} \right.
\end{equation}

For $(k/\kappa)^2 \leq \delta/g$ and according to
Eq.~(\ref{Tbnpone}), the atom cannot leave the cavity in the state
$|b\rangle$. The system becomes in this case very similar to the
elementary problem of the transmission of a structureless particle
through a potential well defined by the cavity and the resonance
amplitudes reach the value 1.

\begin{figure*}[htb]
\includegraphics[width=8.5cm, bb=113 284 475 525]{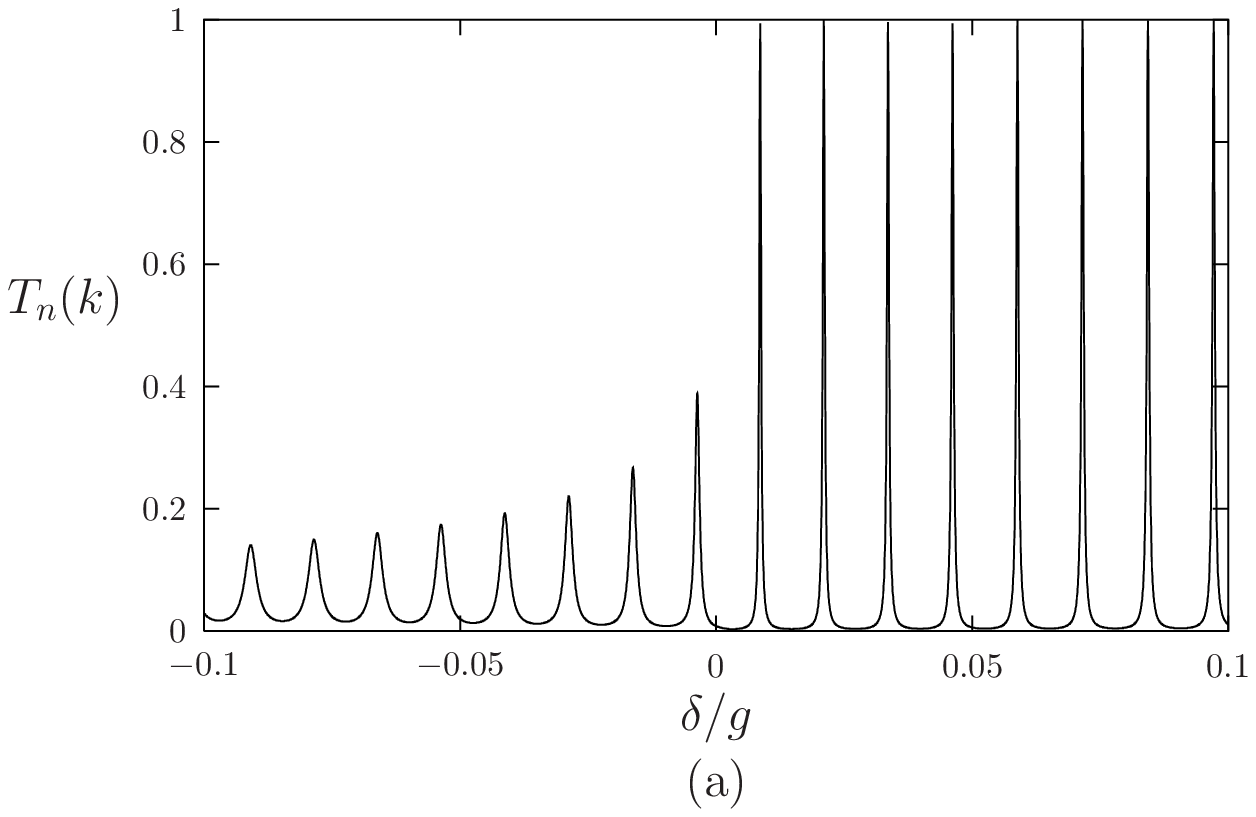}\hspace{5pt}
\includegraphics[width=8.5cm, bb=113 284 475 525]{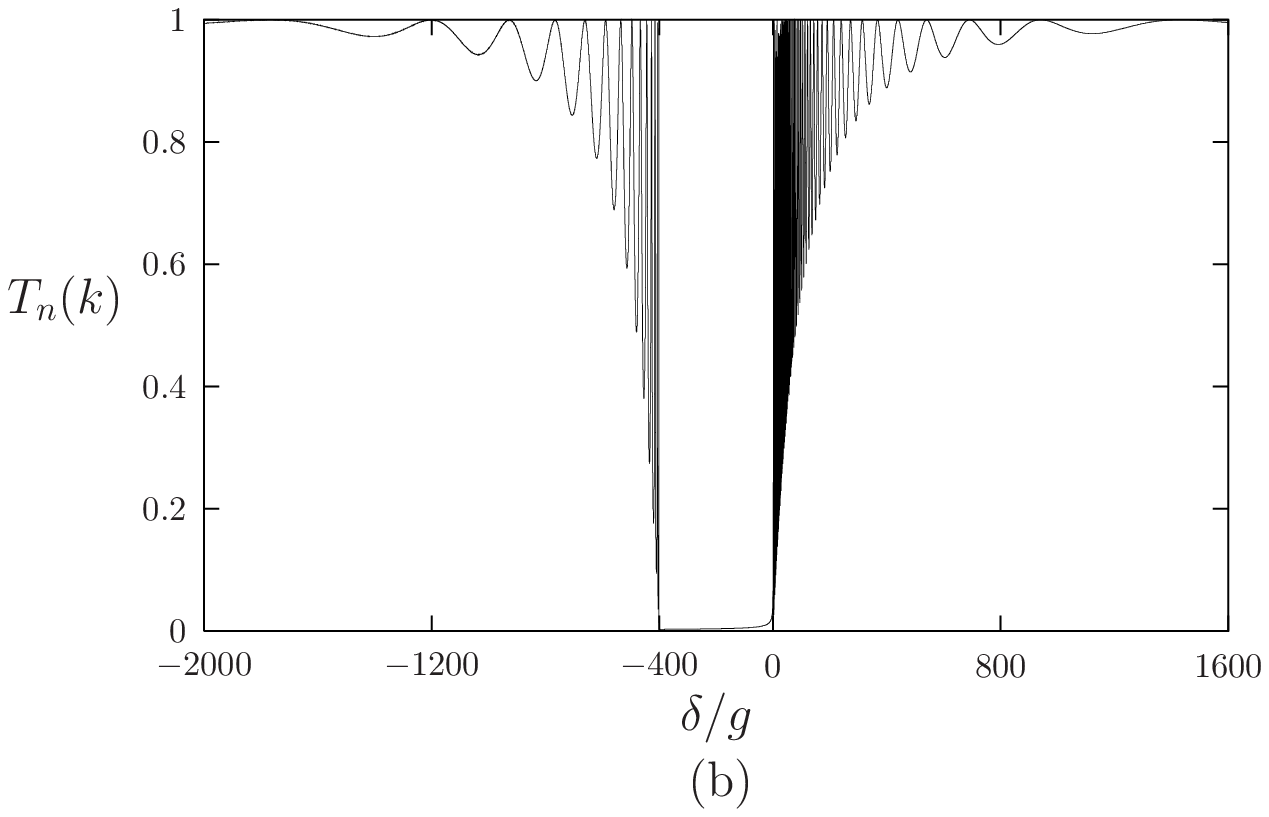}
\caption{Transmission probability of an excited atom through the
mazer with respect to the detuning ($k/\kappa = 0.05$, $\kappa L =
1000$, $n = 0$). The only difference between the two curves
resides in the scale used for the horizontal axis.}
\label{TdsgFig}
\end{figure*}

Same kind of resonances for the transmission probability are
observed with respect to the detuning (see fig.~\ref{TdsgFig}(a)).
For realistic experimental parameters (see discussion in
\cite{Lof97}), these resonances may even become extremely narrow.
Their width amounts only $10^{-2}$~Hz for $\kappa L = 10^5$, $g =
100$ kHz and $k/\kappa = 0.01$. This could define very useful
metrology devices (atomic clocks for example) based on a single
cavity passage and with better performances than what is usually
obtained in the well known Ramsey configuration with two cavities
or two passages through the same cavity \cite{Cla91}.

The same curve of the transmission probability is shown on
fig.~\ref{TdsgFig}(b) over an extended scale. As expected, for
very large (positive or negative) detunings, the atom-cavity
coupling vanishes and the transmission probability tends towards
1. For large positive detunings, $\theta_n \rightarrow \pi/2$ and
this behavior is well predicted by Eq.~(\ref{TransmissionT}) which
yields $T_n(k) \rightarrow 1$. For large negative detunings,
$\theta_n \rightarrow 0$ and we get from Eq.~(\ref{TransmissionT})
$T_n(k) \rightarrow 0$. In fact, when increasing the detuning
towards negative values, the system leaves the cold atom regime
and switches to the hot atom one. For $k \ll \kappa$, this occurs
at the detuning value (see \cite{Bas03b})
\begin{equation}
    -\frac{\delta}{g} = (n + 1) \left( \frac{\kappa}{k}
    \right)^2.
\end{equation}

For large negative detunings, Eq.~(\ref{TransmissionT}) is
therefore no more valid and the transmission probability must be
computed directly using Eqs.~(\ref{Tan}) and (\ref{Tbnpone}). This
explains why the transmission probability changes abruptly at
$\delta/g~\simeq~-400$ on fig.~\ref{TdsgFig}(b), defining this way
a well-defined ``window'' where the transmission probability drops
to a negligible value. This window is all the larger since the
atoms are initially colder.

\section{Velocity selection}

If we consider an atomic beam characterized with a velocity
distribution $\mathcal{P}_i(k)$, each atom will be transmitted
through the cavity with more or less efficiency depending on the
$T_n(k)$ value. The interaction of these atoms with the cavity
will lead through the photon emission process to a progressive
grow of the cavity photon number. By taking into account the
presence of thermal photons and the cavity field damping, Meyer
\etal \cite{Mey97} have shown that a stationary photon
distribution $\mathcal{P}_{\textrm{st}}(n)$ is established inside
the cavity. This distribution is given by
\begin{equation}
\mathcal{P}_{\textrm{st}}(n)=\mathcal{P}_{\textrm{st}}(0)\prod_{m=1}^n\frac{n_b+[r/C]
\overline{\mathcal{P}}_{\textrm{em}}(m-1)/m}{n_b+1},
\end{equation}
where $n_b$ is the mean thermal photon number, $r$ is the atomic
injection rate, $C$ is the cavity loss rate and
$\overline{\mathcal{P}}_{\textrm{em}}(m-1)$ is the mean induced
emission probability
\begin{equation}
\overline{\mathcal{P}}_{\textrm{em}}(n)=\int_0^{\infty}\mathcal{P}_{\textrm{em}}(n,k)\mathcal{P}_i(k)dk,
\end{equation}
with $\mathcal{P}_{\textrm{em}}(n,k)$ the induced emission
probability of a single atom with momentum $\hbar k$ interacting
with the cavity field containing $n$ photons. In presence of a
detuning and in the ultracold regime we have shown in
\cite{Bas03b} that this probability is given by
\begin{equation}
    \label{Pemcold}
    \mathcal{P}_{\textrm{em}}(n,k) = \frac{k_b}{k}\frac{\mathcal{I}(L)}{2} \frac{1 + \frac{\cot \theta_n}{2} \sin(2 \kappa_n \sqrt{\cot \theta_n} L)}{1 + \left( \frac{\kappa_n}{2 k} \right)^2 \cot \theta_n \sin^2(\kappa_n \sqrt{\cot \theta_n}
    L)}.
\end{equation}

After the stationary photon number distribution has been
established, the atomic transmission probabilities in the
$|a\rangle$ state and the $|b\rangle$ state are respectively given
by
\begin{equation}
    T^a(k) = \sum_{n=0}^{\infty} \mathcal{P}_{\textrm{st}}(n)
    T^a_n(k),
\end{equation}
and
\begin{equation}
T^b(k) = \sum_{n=0}^{\infty} \mathcal{P}_{\textrm{st}}(n)
    T^b_{n+1}(k).
\end{equation}

This results in the following final velocity distribution of the
transmitted atomic beam~:
\begin{equation}
    \label{Pfk}
    \mathcal{P}_f(k) = \left\{ \begin{array}{ll}
    P_i(k) T^a(k) + P_i(k')
    T^b(k') & \textrm{ if  } \left(\frac{k}{\kappa}\right)^2 >
    -\frac{\delta}{g},
    \vspace{0.2cm} \\ P_i(k) T^a(k) & \textrm{ otherwise},
    \end{array} \right.
\end{equation}
where $k'$ is such that
\begin{equation}
    k'_b \equiv \sqrt{k'^2 - \kappa^2 \frac{\delta}{g}} = k
\end{equation}
that is
\begin{equation}
    k'^2 = k^2 + \kappa^2 \frac{\delta}{g}
\end{equation}

\begin{figure*}[htb]
\includegraphics[width=8.5cm, bb=113 284 475 525]{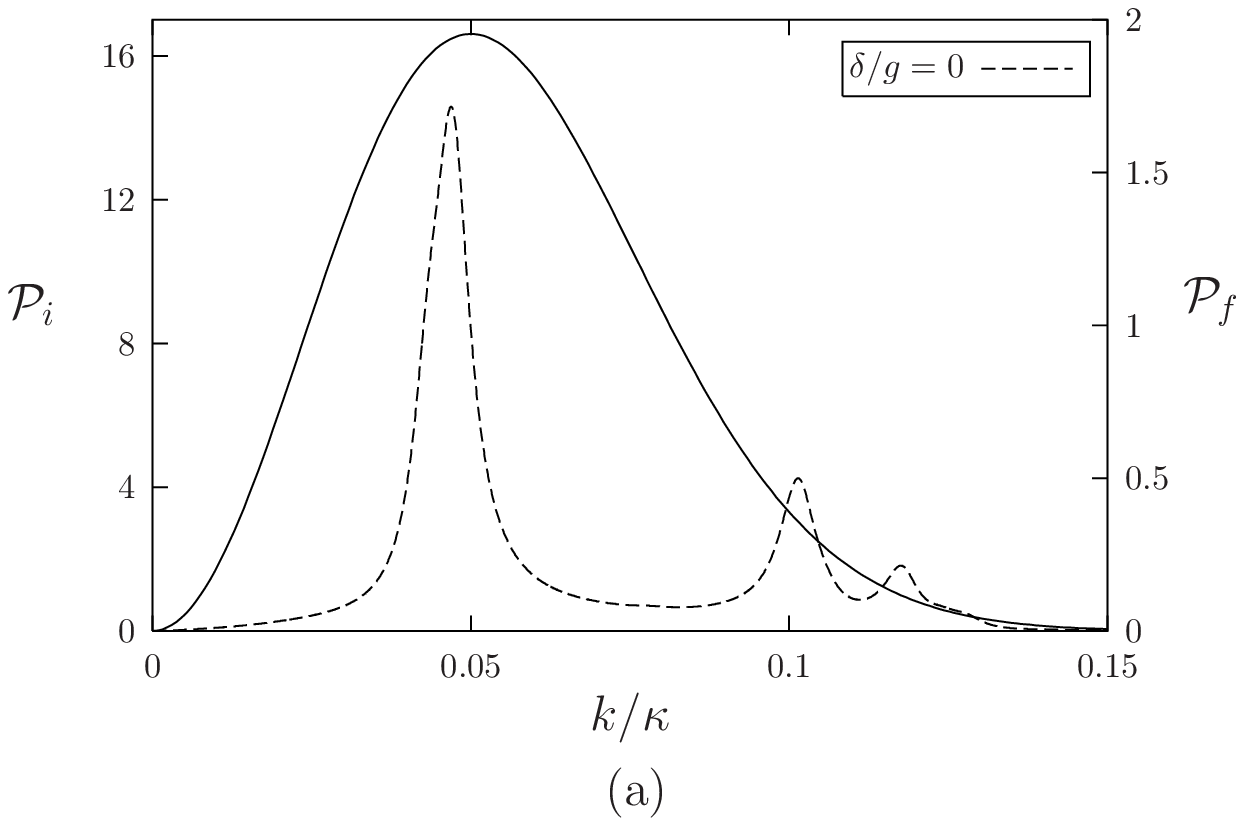}\hspace{5pt}
\includegraphics[width=8.5cm, bb=113 284 475 525]{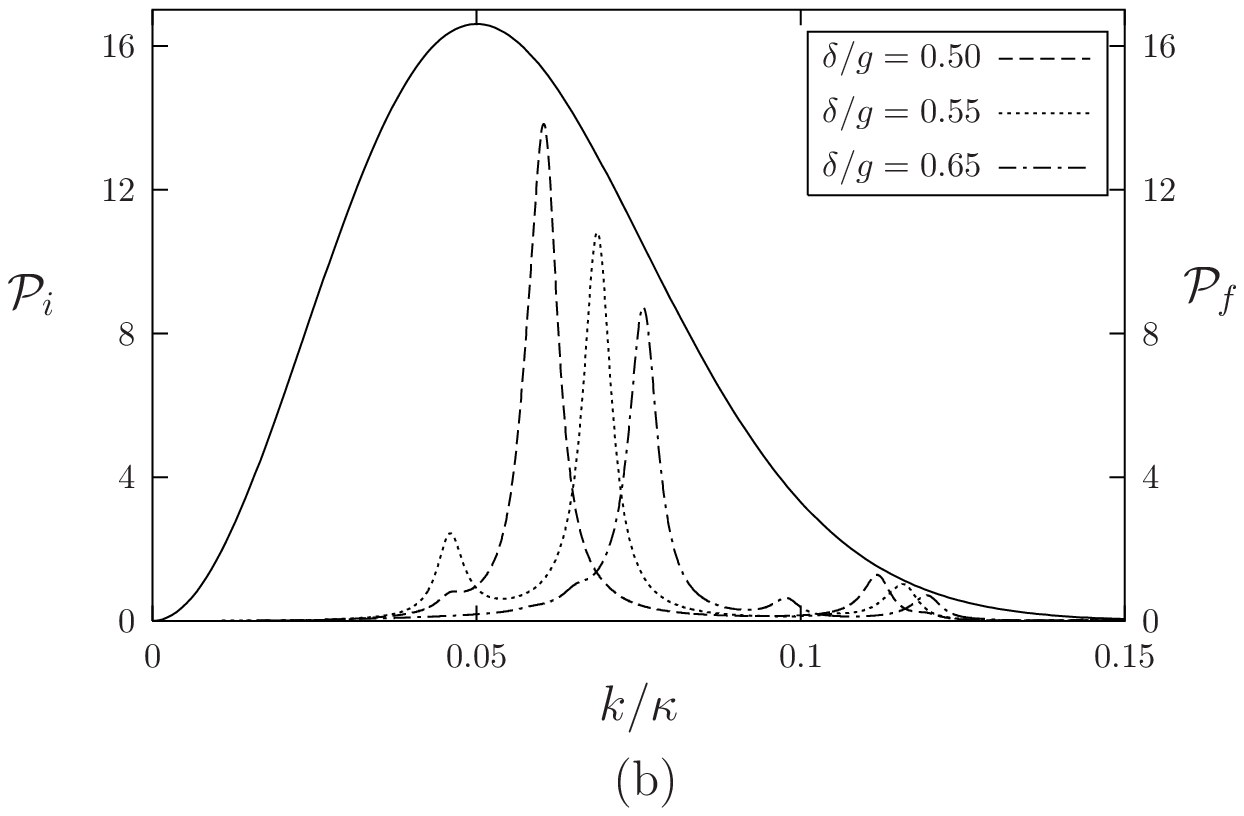}
\caption{Initial (plain curve) and final (dashed curves) velocity
distributions (a) at resonance and (b) for various detuning
values. These curves were computed for $\kappa L = 200 \pi$, $r/C
= 100$ and $n_b = 0.2$.} \label{PifFigs}
\end{figure*}

We show on figs.~\ref{PifFigs} how a Maxwell-Boltzmann
distribution (with $k_0/\kappa = 0.05$ where $k_0$ is the most
probable wave number) is affected when the atoms are sent through
the cavity. The cavity parameters have been taken identical to
those considered in \cite{Lof98} to underline the detuning
effects. We see from these figures that the final distributions
are dominated by a narrow single peak whose position depends
significantly on the detuning value. This could define a very
convenient way to select any desired velocity from an initial
broad distribution. Also, notice from the $\mathcal{P}_f$ scale
that a positive detuning significantly enhances the selection
process. Such detunings indeed maximize the resonances of the
transmission probability through the cavity (see
fig.~\ref{TdsgFig}(a)).

\section{Summary}
In this paper, we have presented the general properties of the
transmission probability of ultracold atoms through a micromaser
in the general off-resonant case. An analytical expression of this
probability has been given in the special case of the mesa mode
function. Particularly, we have shown that this probability
exhibits with respect to the detuning very fine resonances that
could be very useful for metrology devices. We have also
demonstrated that the velocity selection in an atomic beam may be
significantly enhanced and easily tuned by use of a positive
detuning.

\begin{acknowledgments}
\label{Acknowledgments} This work has been supported by the
Belgian Institut Interuniversitaire des Sciences Nucl\'eaires
(IISN). T.~B. wants to thank H. Walther and E. Solano for the
hospitality at Max-Planck-Institut f\"ur Quantenoptik in Garching
(Germany). \end{acknowledgments}

\end{document}